# Cybersecurity in Smart Farming: Canada Market Research

---

By:

**Dr. Ali Dehghantanha**
Canada Research Chair in Cybersecurity (Tier2),
Cyber Science Lab, University of Guelph, Ontario, Canada
EM: ali@cybersciencelab.org

**Dr. Hadis Karimipour**
Director Smart Cyber Physical System Lab
University of Guelph, Ontario, Canada
EM: hkarimi@uoguelph.ca

**Amin Azmoodeh**
PhD Student
Cyber Science Lab, University of Guelph, Ontario, Canada
EM:aazmoode@uoguelph.ca

# Table of Contents

# 1.0 EXECUTIVE SUMMARY

### Introduction & Methodology

The Cyber Science Lab (CSL) and Smart Cyber-Physical System (SCPS) Lab at the University of Guelph conduct a market study of cybersecurity technology adoption and requirements for smart and precision farming in Canada. We conducted 17 stakeholder/key opinion leader interviews in Canada and the USA, as well as conducting extensive secondary research, to complete this study. Each interview generally required 15-20 minutes to complete. Interviews were conducted using a client-approved interview guide. Secondary and primary research focussed on the following areas of investigation:

- Market size and segmentation
- Market forecast and growth rate
- Competitive landscape
- Market challenges/barriers to entry
- Market trends/growth drivers
- Adoption/commercialization of the technology

### Conclusions

Following are the conclusions of this report:

### Market Data – Secondary Research

1. **Smart agriculture and cybersecurity are both large and rapidly growing markets individually**. As of 2019, the global smart agriculture market is valued at $10.3 billion, with a CAGR of 14.1%. The global cybersecurity market is even larger, at $145 billion, with a CAGR of 10%.

2. **But currently (as of Aug 2020), cybersecurity in smart agriculture is a relatively small market**. Smart/digital agriculture is one of the smaller cybersecurity markets, however, accounting for approximately 2-3% of the total spend in 2019, or $1.1-1.6 billion in North America.

3. Secondary sources indicate that **cybersecurity in smart agriculture is still several years away from hitting critical mass**, and lags other sectors such as defense/government, banking, and finance, healthcare, IT and telecommunications, automotive and retail. (and this opinion was corroborated in stakeholder interviews).
4. **Market trends and growth drivers** for digital agriculture/smart farming are expected to be - industry investment in hardware by the major players, increased productivity, greater efficiency of field inputs, and less labour intensiveness.

5. **Market challenges/barriers to entry** for digital agriculture/smart farming are expected to be – cost, lack of standardization, reluctance to change, and lack of broadband infrastructure.

### Stakeholder Interviews – Reaction to the Client's Technology Value Proposition

#### Cybersecurity, Manufacturing and Government Stakeholders

6. There is currently a **<u>moderate</u> need (6.6/10)** for cybersecurity systems for protecting smart and precision farming in Canada.

7. **<u>Cost</u> was mentioned most often as the most significant challenge/barrier to entry**, followed by the need for the simplicity of systems, rural connectivity issues, understanding of the problem, and lack of agreement as to standards.

#### Digital Agriculture /Farming Stakeholders

8. **Cybersecurity is a <u>moderately significant</u> issue (6.1/10)** to digital agriculture/farming stakeholders.
9. **<u>No</u> cyberattacks of data breaches have been experienced** by the digital agriculture / farming stakeholders interviewed.
10. Digital agriculture / farming stakeholders are **generally <u>not satisfied</u> with existing cybersecurity solutions** – rating existing solutions only 4.5/10 in terms of satisfaction.

## Technology Specialist Conclusions

11. **There is a customer market need** for the technology value proposition**, but it is just emerging**.

12. A company commercializing a cybersecurity solution in smart-ag will have to **overcome several technology and marketing challenges in their "go to market strategy"** in order to be financially sustainable.

    a) There will be **non-traditional technology competitors to an agri-food specialized cybersecurity services company, and they will be formidable**.

    b) A new organization may also choose to **research and educate the industry** on overlapping technologies such as blockchain and machine learning.

    c) One implication of the globalization of cybersecurity services will be that **it is unlikely that many of the servers and technology will be based in Canada**, so there may be reasons for a "Made in Canada" services offering – e.g., the Cannabis industry needs to avoid US legal restrictions.

    d) There is a need for a **significant investment in stakeholder education** to ensure our agriculture and food sector is doing everything it can from a policy perspective.

## 2.2 Objectives - Market Research Task Required

The goal of this study is to identify the potential market opportunity for **Cybersecurity Technologies in Protecting Smart & Precision Farming** environment. This will provide critical guidance for the next stages of technical development, and refine and better develop the commercialization strategy for cybersecurity solutions in the Ag space. Therefore, key questions to be answered include:

**(1) Market size and forecast – North America** (USA & Canada) **& Global**

- Current market size estimate for smart farming data transmission/storage and cybersecurity solutions for this space
- The focus will be primarily on Canada and the USA, and secondarily on the global market
- Market Forecast (5-year forecast) and CAGR (Compound Annual Growth Rate)

**(2) Competitive Landscape**

- What are satisfaction levels with currently available technologies, and why?
- What new, relevant, competitive developments are currently underway?

**(3) Market Challenges/Barriers to Entry**

- Identification of the main challenges/barriers.
- What issues must be addressed, or obstacles overcome to gain market adoption. (e.g., competitive pressures, certifications, approvals, cost constraints, etc.)

**(4) Market Trends and Growth Drivers**

**(5) Adoption/Commercialization of the Technology**

- Who would most likely use and/or purchase this technology? Who would most likely influence the usage and/or purchase decision?
- How much do customers currently pay for alternative/substitute products?
- How likely are they to switch to or adopt this technology? Depending on what factors?
- How much would they be willing to pay for this technology?

**(6) Product Positioning**

- What features would users like to see in this technology?
- What elements of the value proposition resonate most strongly with stakeholders/have the most potential value? In which sectors?
- How is the proposed solution perceived as different from other technologies, and why would someone choose the proposed solution?

## 2.3 Study Methodology

### Project Orientation

This project began with an orientation session held with the cybersecurity technology expert representatives of the CSL and SCPS Lab - University of Guelph. The technology expert was given an opportunity to discuss the technology and collect any relevant materials, data/reports that the clients could provide that would assist the study. A key output of the orientation session was to begin to develop a clearly articulated value proposition that will be shareable with stakeholder interviewees.

### Secondary Market Data Review & Analysis

We conducted a referenced literature review and analysis of the relevant market spaces, including market data reported for market size and growth estimates, market trends and profile information on key players in this space (major end-users, major Canadian and US manufacturers, distributors, associations, as well as applicable regulatory agencies and industry observers).

### Primary Market Research – Key Stakeholder Interviews

Market stakeholder interviews are key to the successful execution of this project. We conducted in-depth interviews by telephone, with a total of **17 key stakeholders**. Specific interviews were with:

- **Technology end-users** – smart farming/agriculture
- **Cybersecurity** providers and experts
- **Industry observers –** including academic and relevant government agencies, as well as applicable industry/trade associations.
- **Additional industry contacts** identified through secondary research or within the Technical Advisor or Client's existing network.

Each of these interviews generally required 15-20 minutes. Interviews were executed using a **formal, client-approved interview guide** that focussed on the following key directions of investigation.

- Satisfaction with existing solutions/current technologies
- The size/significance of the problem the proposed technology is looking to solve
- Competitive Landscape
- Market Challenges/Barriers to Entry
- Market Trends and Growth Drivers
- Reaction to the Technology Value Proposition
- Adoption of the Technology
- Product Positioning

### Analysis & Reporting

All information was collected, analyzed, and this written report provided. Data collected from secondary and primary sources were analyzed to determine the attractiveness of the market opportunity and commercial potential of cybersecurity technologies based on key market factors (size and growth rates, competitive intensity, paths to the market entry) and response to the technology value proposition.

# 2.0 RESULTS – PUBLISHED DATA

The adoption of Agricultural Technology (**agtech**) is changing the agricultural industry in many ways, and the terminology is adapting as well. There's ***Smart Farming*** (SF), ***Precision Agriculture*** (PA) and ***Digital Farming*** (DF), which is itself an amalgam of SF and PA. The ***Agrocares***[1] site helps sort out some of it for us.

**Precision farming** and precision agriculture are generally regarded as the same thing. However, the term **precision agriculture**, often abbreviated **PA**, is more widely used. In a 2016 report on how big data will revolutionize the global food chain, *McKinsey & Company* define **precision agriculture** as: "*a technology-enabled approach to farming management that observes, measures, and analyses the needs of individual fields and crops*". According to McKinsey, the development of precision agriculture is shaped by two trends: "*big-data and advanced-analytics capabilities on the one hand, and robotics—aerial imagery, sensors, sophisticated local weather forecasts—on the other*".

***Smart farming*** *involves not just individual machines but all farm operations. Farmers can use mobile devices such as smart phones and tablets to access real-time data about the condition of soil and plants, terrain, climate, weather, resource usage, manpower, funding, etc. As a result, farmers have the information needed to make informed decisions based on concrete data, rather than their intuition.*

*The essence of* ***digital farming*** *lies in creating value from data. Digital Farming means to go beyond the mere presence and availability of data and create actionable intelligence and meaningful added value from such data.*

*Digital farming is integrating both concepts - precision farming and smart farming. According to a paper on Digital Agriculture by DLG*[2] *(German Agricultural Society), digital farming is understood to mean "consistent application of the methods of precision farming and smart farming, internal and external networking of the farm and use of web-based data platforms together with Big Data analyses"*

Wikipedia[3] expands on the technology that makes Precision Agriculture possible: **The practice of precision agriculture has been enabled by the advent of GPS and GNSS[4]**. The farmer's and/or researcher's ability to locate their precise position in a field allows for the creation of maps of the spatial variability of as many variables as can be measured (e.g. crop yield, terrain features/topography, organic matter content, moisture levels, nitrogen levels, pH, EC, Mg, K, and others). Similar data is collected by sensor arrays mounted on GPS-equipped combine harvesters. These arrays consist of real-time sensors that measure everything from chlorophyll levels to plant water status, along with multispectral imagery. This data is used in conjunction with satellite imagery by variable rate technology (VRT) including seeders, sprayers, etc. to optimally distribute resources.

Precision agriculture has also been enabled by unmanned aerial vehicles-drones, which are relatively inexpensive and can be operated by novice pilots. These agricultural drones can be equipped with hyperspectral or RGB[5] cameras to capture many images of a field that can be processed using photogrammetric methods to create orthophotos and NDVI[6] maps. These drones can capture several metric points of the land that can later be used to deliver proper water and fertilization to crops.

A *Sciforce*[7] article on the IOTForAll site describes some of the new tools in the modern Farmers' toolbox:

---

[1]Agrocares, **What is the difference between precision, digital and smart farming?** , https://www.agrocares.com/en/news/precision-digital-smart-farming/

[2] CEMA, **Digital Farming: what does it really mean?**, https://www.cema-agri.org/images/publications/position-papers/CEMA_Digital_Farming_-_Agriculture_4.0__13_02_2017_0.pdf

[3]Wikipedia, **Precision agriculture**, https://en.wikipedia.org/wiki/Precision_agriculture

[4] GNSS = Global Navigation Satellite System

[5] RGB = Red Green Blue - A *camera* equipped with a standard CMOS sensor through which the colored images of persons and objects are acquired

[6]The normalized difference vegetation index (NDVI) is a simple graphical indicator that can be used to analyze remote sensing measurements, typically, but not necessarily, from a space platform, and assess whether the target being observed contains live green vegetation or not.

[7] Sciforce, **Smart Farming: The Future of Agriculture**, https://www.iotforall.com/smart-farming-future-of-agriculture/

Among the technologies available for present-day farmers are:

- **Sensors**: soil, water, light, humidity, temperature management
- **Software**: specialized software solutions that target specific farm types or use case agnostic IoT platforms
- **Connectivity**: cellular, LoRa, *etc*.
- **Location**: GPS, Satellite, *etc*.
- **Robotics**: Autonomous tractors, processing facilities, *etc*.
- **Data analytics**: standalone analytics solutions, data pipelines for downstream solutions, *etc*.

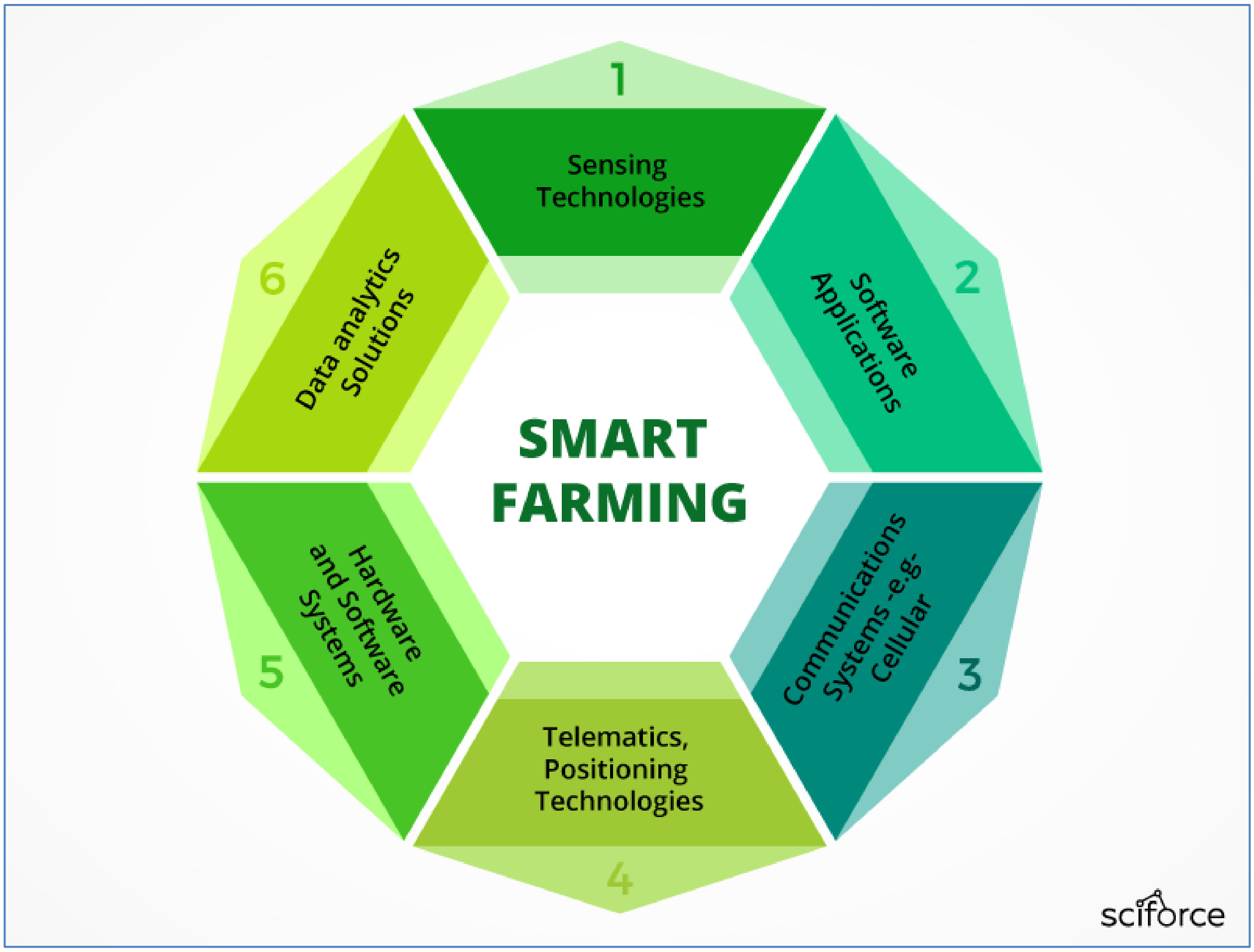

*Source: Sciforce, IoTforall.com (2019)*

## 2.1 Cybersecurity in Digital Agriculture /Smart Farming

### Definitions

Although "Smart Farming" and "Precision Agriculture" have been used to describe this segment, it appears that the term **Smart Agriculture** is being increasingly adopted as a more inclusive terminology.

## Recent Trends - Background

According to a recent whitepaper by the *Jahn Research Group*[8] at the *University of Wisconsin-Madison* entitled 'Cyber Risk and Security Implications in Smart Agriculture and Food Systems', rapid changes in American agriculture and how food is produced and distributed are opening new and often unappreciated cyber attack vectors with unappreciated economic and security implications. The structure and operation of modern highly "networked" food systems (and the obvious requirement for functional energy, transportation, and other systems) fundamentally depend on networked information systems, some of which may not be secured from cyber attacks. The same vulnerabilities also make food systems highly vulnerable to hybrid warfare tactics of both state and non-state actors. The combined complexities of these networked systems interacting together stand to amplify threats and vulnerabilities that exist in any of the major systems, as well as risk to other dependent systems. The result is uncharacterized risks that are highly relevant for food safety and supply, manufacturing, banking, financial, commodities, insurance, and other sectors.

Even a short-duration interruption in the refrigeration chain or other essential infrastructure for food distribution, or a targeted disruption of a highly time-sensitive process such as harvest, could cause major, long-lasting effects globally and significant economic losses. In fact, past cyber events that were neither well-timed nor coordinated have caused the mass disruption, e.g., disruption of markets in the Sony attack, while well-coordinated attacks, usually attributed to state actors (*Stuxnet/Saudi Aramko/Russia Ukraine Power*), could also be devastating. If the actor was trying to build a profile (usually lone actor) or simply vandalize (i.e. college hackers), it is not inconceivable given the potential vulnerabilities we highlight below that the attack could be "lucky" and cause real damage with cascading effects throughout the system. The authors **conclude that competitor-on-competitor attacks also cannot be ruled out in this sector, especially given the global nature of supply chains.** In addition to this and other similar direct effects of cyber-insecurity on food systems, there are a host of other indirect and secondary impacts that could negatively affect global and national security.

Because of this interconnectedness and the increasing application of smart technology and devices, the risk of the American agricultural industry being negatively impacted by a service interruption caused by a **cyberattack or accidents, acts of nature or AI/autonomous systems (collectively "AAAA Threats") is rapidly growing**. The exposure is a result of a failure of education and market information, since the issue is not yet well known or understood by equipment manufacturers or producers, **and equipment consumers are <u>not yet</u> demanding that the equipment they purchase be cyber secure**. This leaves not just North Americans but all consumers across the globe vulnerable to price shocks or shortages resulting from a cyber attack in North America.

In 2018, the *US Council of Economic Advisers* reported the agricultural sector experienced 11 cyber incidents in 2016. Compared to other sectors, such as transportation or manufacturing, **the agricultural sector experienced <u>a relatively low number</u> of reported cyber incidents**. While historical data show lower "likelihoods" of such attacks in the agricultural sector, the externalities of insufficient cyber protection, spillovers of attacks on linked sectors, and the growing implementation of cyber devices in general and in the agricultural sector, in particular, suggests that the severity of any such incident or attack could be more profound in the near future. Cyber attacks such as the 2017 WannaCry ransomware and Petya malware illustrate the potential danger to American agriculture as smart technology is increasingly deployed. Operating systems in many countries were compromised as the ransomware and malware took control of internet-dependent operating systems that had not been properly updated with patches.38 WannaCry victims, for example, found that files were encrypted and payment of a ransom of $300 in bitcoins was demanded, with the payment demand doubling after three days.

[8] Jahn Research Group, **Cyber Risk and Security Implications in Smart Agriculture and Food Systems**, https://jahnresearchgroup.cals.wisc.edu/wp-content/uploads/sites/223/2019/01/Agricultural-Cyber-Risk-and-Security.pdf, 2019

The ISA[9] website warns against complacency in cybersecurity for SF and PA in particular: All industries are susceptible to cyberattacks, but agriculture is especially vulnerable. Newly networked precision farming technologies have taken a highly mechanical and labour-intensive industry online, significantly increasing the threat landscape. Otherwise common threat vectors like the improper use of USB drives or spear-phishing may have unique and far-reaching consequences on the agricultural industry, and, correspondingly, negatively impact the food supply chain.

*David Kohl, Professor Emeritus of Agricultural Finance, Virginia Tech University*, says, "The more high-tech we become, the more vulnerable we become." **Cyber attacks on agricultural technology could be a black swan facing the agricultural industry over the next decade**. Kohl goes on to state, "It's not if it's going to happen; it's when it's going to happen." Magnifying the agricultural industry's susceptibility is the fact that much of the technology that precision farming now relies on are smart devices that are part of the Internet of Things (IoT). IoT devices have primarily been produced with few internal cybersecurity measures in place.

At the 2018 *Grain World Conference* in Winnipeg, *Doug Tait*, cybersecurity legal expert at *Thompson, Dorfman, Sweatman*, spoke about the importance of cybersecurity on farms. Tait said the myriad of IoT devices being used in agriculture are creating threats that most farmers haven't considered. "The agriculture sector alone is experiencing an IOT growth rate of 20 percent compounded yearly, and by 2025 there will be an estimated 224 million devices," Tait said.

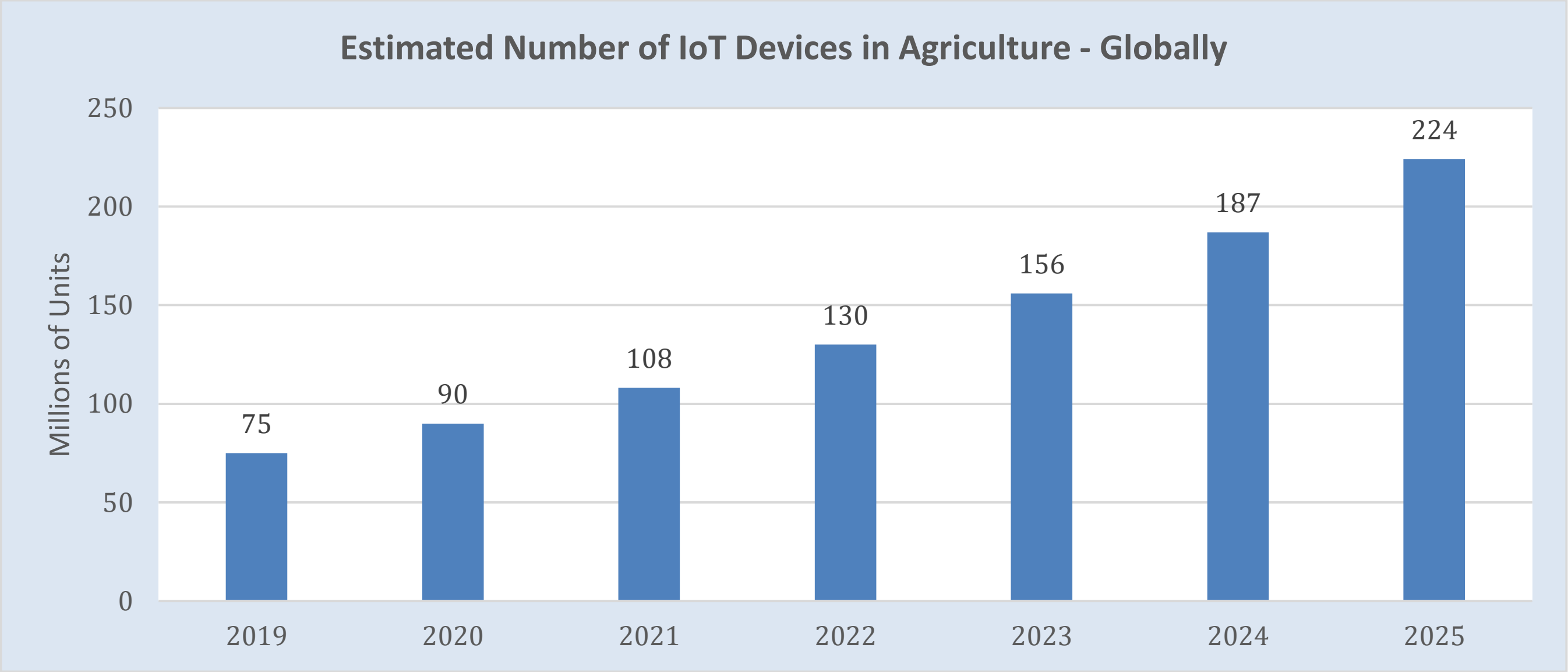

*Source: Thompson, Dorfman, Sweatman (2018)*

[9]ISA, **Cybersecurity for the Agricultural Industry**, https://www.e-isa.com/blog/cybersecurity-for-agriculture.html

### 2.1.1 Market Size, Segmentation & Trends

In this section of the report, we characterize the size of the smart agriculture market, and then review the cybersecurity market, focusing on the relative importance of cybersecurity in smart agriculture in comparison to other industries.

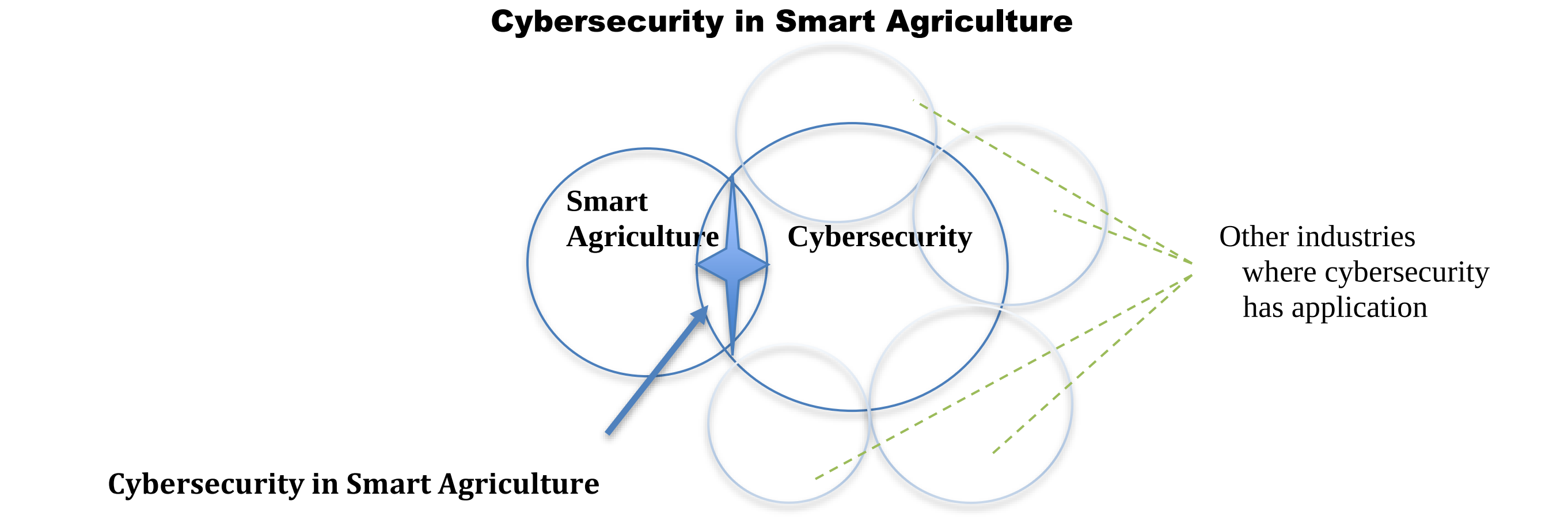

#### a) Smart Agriculture

According to a report from *Variant Market Research*[10], the **Global Smart Agriculture Market** will reach $20 billion by 2024; growing at a CAGR of 14.1% from 2016 to 2024.

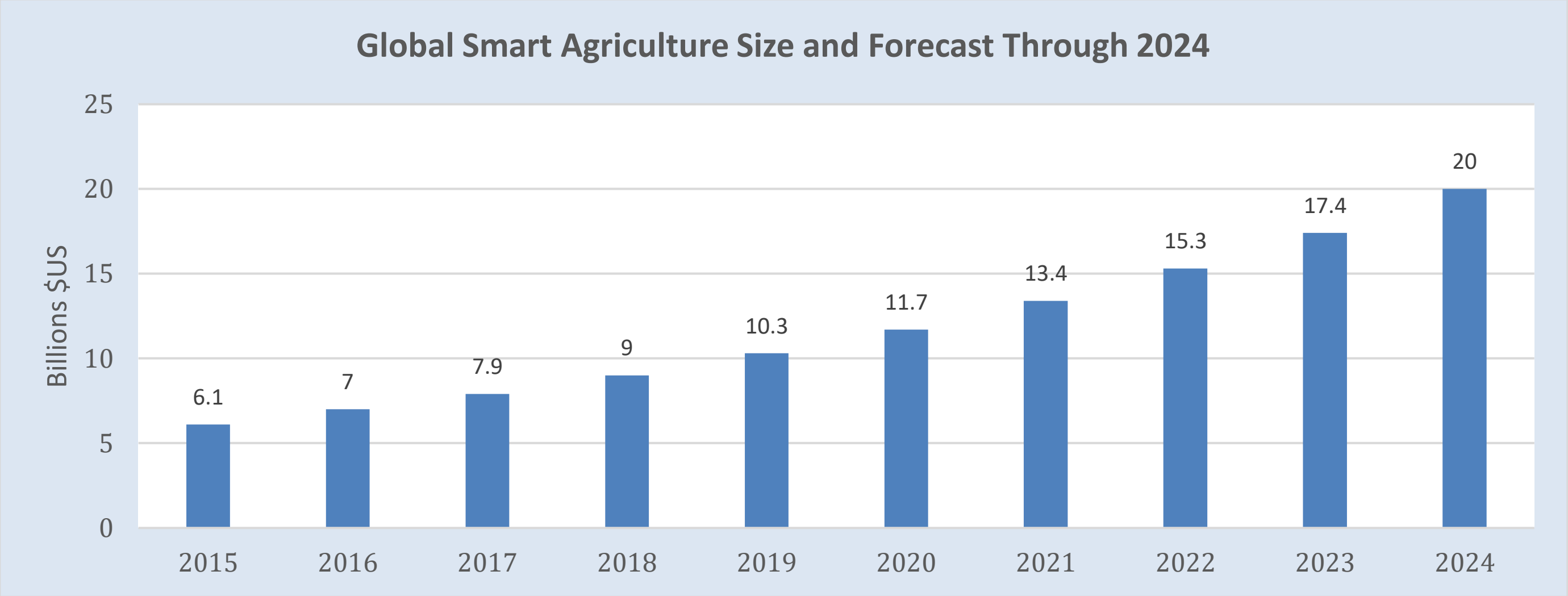

*Source: Variant Market Research (2016)*

The global **smart agriculture** market can be categorized by solutions, application and geography. Solution segments include agriculture asset management, network management, supervisory control & data acquisition, smart water management, logistics & supply chain management, and others. Application segments include livestock monitoring, precision agriculture, fish farming, smart greenhouse and others.

#### b) Cybersecurity

[10] Variant Market Research, **Smart Agriculture Market (By Solution: Agriculture Asset Management, Network Management, Supervisory Control & Data Acquisition, Smart Water Management Logistics & Supply Chain Management, and Others; By Application: Livestock Monitoring, Precision Agriculture, Smart Greenhouse, Fish Farming, and Others; By Geography: North America, Europe, Asia-Pacific and RoW) Global Scenario, Market Size, Outlook, Trend and Forecast, 2015-2024** https://www.variantmarketresearch.com/report-categories/information-communication-technology/smart-agriculture-market

According to *Steve Morgan, the editor in chief of Cybersecurity Ventures*[11], "In 2004, the global cybersecurity market was worth $3.5 billion — and in 2017 it was expected to be worth more than $120 billion. The cybersecurity market grew by roughly 35X over 13 years, entering our most recent prediction cycle. Cybersecurity Ventures predicts global spending on cybersecurity products and services will exceed $1 trillion cumulatively over the five years from 2017 to 2021."

"While all other tech sectors are driven by reducing inefficiencies and increasing productivity, cybersecurity spending is driven by cybercrime. The unprecedented cybercriminal activity we are witnessing is generating so much cyber spending, it's become nearly impossible for analysts to accurately track." "**We anticipate 12-15 percent year-over-year cybersecurity market growth through 2021**, compared to the 8-10 percent projected by several industry analysts."[12]

"IT analyst forecasts are unable to keep pace with the dramatic rise in cybercrime, the ransomware epidemic, the refocusing of malware from PCs and laptops to smartphones and mobile devices, the deployment of billions of under-protected Internet of Things (IoT) devices, the legions of hackers-for-hire, and the more sophisticated cyberattacks launching at businesses, governments, educational institutions, and consumers globally."

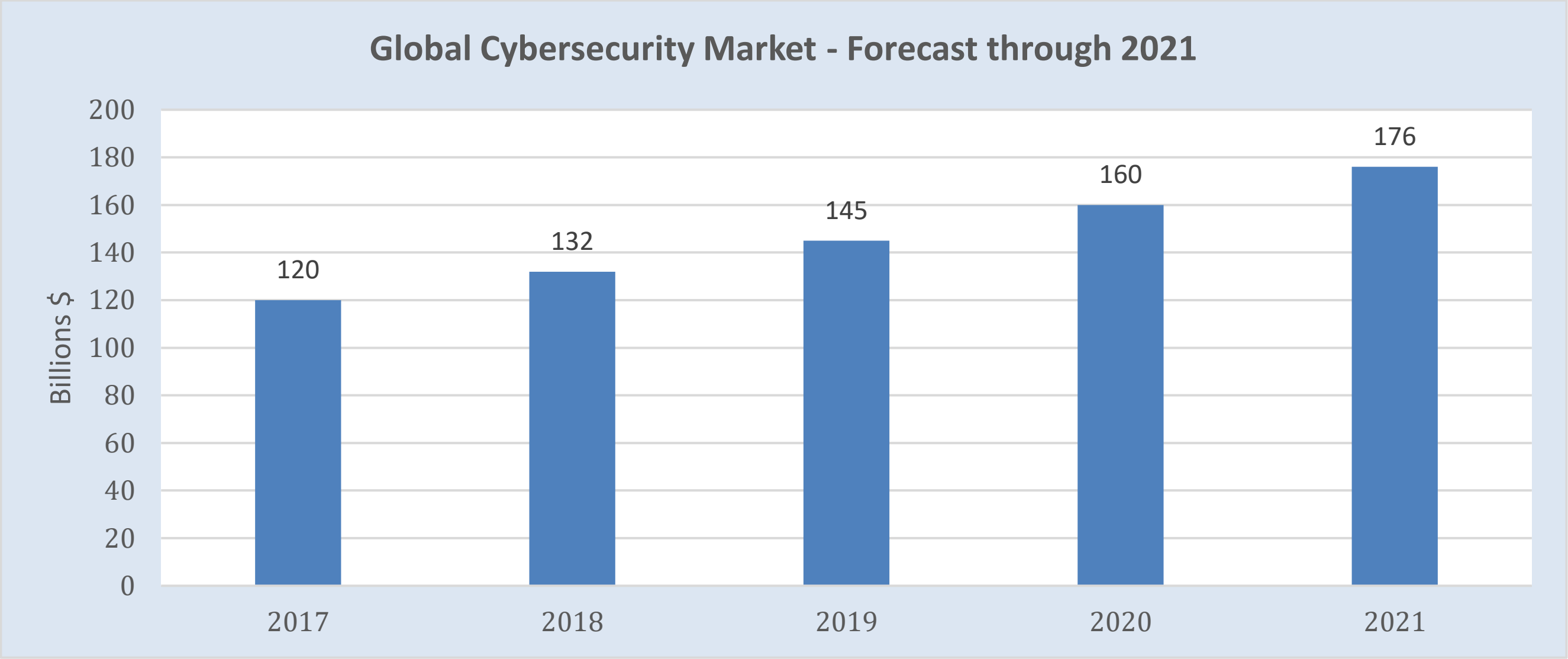

*Sources: Cybersecurity Ventures, Adventus estimates (2019)*

According to *Grandview Research*[13], North America currently accounts for 38% of the global cybersecurity market, or $ 55 billion in 2019.

---

[11] Steve Morgan, **Global Cybersecurity Spending Predicted To Exceed $1 Trillion From 2017-2021**, June, 2019, https://cybersecurityventures.com/cybersecurity-market-report/
[12] The Adventus forecast on the chart below takes a middle estimate of a CAGR of 10%.
[13] Grandview Research, **Cybersecurity Market Size, Share & Trends Analysis Report By Component, By Security Type, By Solution, By Service, By Deployment, By Organization, By Application, And Segment Forecasts, 2019 – 2025,** May 2019, https://www.grandviewresearch.com/industry-analysis/cyber-security-market

### c) Cybersecurity in Smart Agriculture

Although Smart Farming and Precision Agriculture are still in the relatively early stages of technology adoption and integration, the need for investment in cybersecurity is immediate. If it follows other industries, the implementation of cybersecurity in digital agriculture will be recognized first as a counter to fears of cybercrime, and then as a response to actual attempts to disrupt operations and attempt data breaches of unprotected systems. The prediction of up to a quarter billion under-protected Internet of Things (IoT)[14] devices being integrated into digital farming represents a major vulnerability that will need to be addressed[15].

The *US Congressional Budget Office* (CBO)[16] has estimated that implementing the Internet of Things (IoT) Cybersecurity Improvement Act of 2019 will cost US $35 million over the 2019-2024 period, plus an additional US $11 million in 2020 to develop the IoT guidelines and standards mandated in the legislation.

According to *ISA*, the growth of cybersecurity specific to digital agriculture will be driven by both the adoption of technologies new to the industry and the need to safeguard businesses and supply chains. The threats could range from (relatively) straightforward ransomware to agroterrorists[17] seeking to disrupt farming operations through access to automated systems. Even as a subset of the multi-billion-dollar industry that is Cybersecurity, there will be substantial investment required by and for the digital farming community.

According to *Grandview Research*[18], by application the defense/government application segment led the market in 2018 and accounted for almost 27% share of the overall revenue. The growth is attributed to high spending on the security solutions and growing concerns over data breach of intelligence, research and development activities, and financial activities, which can potentially lead to cyber terrorism. Moreover, the government sector is increasingly targeted with attacks including negligent insiders, malicious malware, and zero-day attack. Thus, mitigation of online attacks through robust security architect is a key trend, which will drive the segment over the forecast period. *Grandview* groups digital agriculture into its “other” category, behind banking and finance (21%), healthcare (13%), IT and telecommunications (11%), automotive (10%), and retail (10%) – with “other accounting for 8%. While no breakout exists for this category, **it would be reasonable to assume that <u>digital agriculture</u> could account for between 25-33% of the “other” category, or <u>2-3% of the total cybersecurity market</u>.**

**With a North America cybersecurity market of $ 55 billion in 2019, and a market share estimate of 2-3%, that would make the North American cybersecurity for digital agriculture market worth approximately <u>$1.1-1.6 billion in 2019</u>.**

---

[14] IoT Agenda, **IoT cybersecurity: A major concern for agriculture**, https://internetofthingsagenda.techtarget.com/blog/IoT-Agenda/IoT-cybersecurity-A-major-concern-for-agriculture

[15] IoT Agenda, **IoT security (internet of things security)**, https://internetofthingsagenda.techtarget.com/definition/IoT-security-Internet-of-Things-security

[16] TheInternetofallthings, **Cost of US IoT cybersecurity plan over $35 million** https://www.theinternetofallthings.com/cost-of-us-iot-cybersecurity-plan-over-35-million

[17] ISA, **Cybersecurity for the Agricultural Industry**, https://www.e-isa.com/blog/cybersecurity-for-agriculture.html

[18] Grandview Research, **Cybersecurity Market Size, Share & Trends Analysis Report By Component, By Security Type, By Solution, By Service, By Deployment, By Organization, By Application, And Segment Forecasts, 2019 – 2025,** May 2019, https://www.grandviewresearch.com/industry-analysis/cyber-security-market

### 2.1.2 Market Trends/Growth Drivers

Following are the main growth drivers in digital agriculture/smart farming:

1. **Investment**: Indeed, the hardware distribution players (*Agrium, Simplot, Wilbur Ellis*) and "Big Six" majors are already engaged or have developed their own platforms and are looking for ways to partner with the potential new "disruptors" entering the market. For instance, *John Deere* and *IBM* have been developing smart-farm equipment and integrated software for years and have recently started to foray into web and mobile-based tools. However, now VCs and entrepreneurs have fully turned their attention to technology opportunities in agriculture, with *AgFunder*[19] noting that Agri-food Tech startups, those innovating all the way from farm-to-fork, raised $16.9 billion in 2018, a 43% year-over-year increase, in-line with the overall VC market globally. In contrast to all VC industries, there was an 11% increase in deal activity, particularly at the seed stage where levels have contracted in other industries.
2. **Increased productivity**[20]: The major ag players view the ability to offer new products and services to fill their pipeline as the key to their future growth — and digital ag has the potential to do just that. Digital ag is about improving the decisions farmers make about planting, fertilization, crop protection, and irrigation based on the inputs, as well as yield and quality outputs, at harvest.
3. **More efficient use of field inputs**[21]—particularly nitrogen fertilizer—is essential for several reasons. For one, fertilizer costs are rising. Fertilizer sales now exceed $18 billion annually in the United States and represent between 30–50% of the cost of production for wheat and corn on most farms.
4. **Less labour intensive[22]: "**The vegetable industry hasn't adopted automation in a slow, steady way", Richard Smith, a Vegetable Crop Production and Weed Science Farm Advisor with the University of California, says. Growers were hesitant for years, then in 2011 lettuce growers began embracing auto thinners in a big way. "It was nothing, then a big surge," he says. The turning point, he thinks, was the Southwest Ag Summit, where growers were able to see an auto thinner in action and observe that it worked. The potential for increased productivity and reduced labour costs due to automation helps explain why the implementation of IOT devices is growing by 20% annually.

---

[19] https://agfunder.com/research/agri-food-tech-investing-report-2018/
[20] https://techcrunch.com/2016/08/08/driving-new-revenue-sources-through-a-digital-ag-revolution/
[21] https://www.nap.edu/read/13292/chapter/20#203
[22] https://www.precisionag.com/in-field-technologies/precision-agriculture-in-specialty-crops-labor-cost-savings-are-key/

### 2.1.3 Market Challenges/Barriers to Entry

Following are the main market challenges/barriers to entry in digital agriculture/smart farming:

1. **Cost**[23]: Digital ag success stories will be built on demonstrating utility, the economic gain per acre farmed. Based on accepted industry norms, **farmers expect $3 of gain for every $1 invested**. Many farms are operating with older equipment that is not compatible with newer technology and would require a substantial investment to upgrade or replace.

2. **Lack of standardization**[24]: Despite the increasing adoption rate of agricultural apps by growers, no single application has been widely adopted, and the market is in its infancy.

   One of the major challenges software companies are facing is that, while most startups focus on solving one aspect of the supply chain, growers need app and software solutions for the entire production process. When growers are asked what their main expectation from an agricultural software or app is, most say that it must be **simple and easy to use** and that it must provide a solution for "all" (i.e., for at least the three main practices **– irrigation, crop protection, and fertilization**).

3. **Reluctance to change**: However, the current wave of ag innovation will not necessarily equal strong market adoption. The adoption rate of new technology by farmers is notoriously slow — self-steering tractors and combines, for example, are less than 20% adopted after a decade or more of effort. Farmers demand a step change in value. Indeed, adoption is the key challenge facing these players, new and old alike. Certainly, progressive farmers like *Tom Farms* or *International Farming Corp*. are counterpoints that are aggressive about capturing better yield and sustainability, but they are pathfinders, not the mainstream. It may take the next generation of farmer before adoption becomes widespread.

4. **Lack of broadband infrastructure**[25]**:** What's abundantly clear from the report is that while cost is always top of mind for farmers, in Western Canada, it's actually **Internet access and speed** that ranks as a top barrier, followed closely by a lack of knowledgeable people involved.

---

[23] https://techcrunch.com/2016/08/08/driving-new-revenue-sources-through-a-digital-ag-revolu
[24] https://www.precisionag.com/digital-farming/analytics/why-are-570-million-farmers-not-yet-using-agricultural-apps/
[25] https://www.realagriculture.com/2017/05/whats-holding-precision-farming-back/

# 3.0 RESULTS - STAKEHOLDER INTERVIEWS

Interviews were held with the following 17 stakeholders (14 in Canada and 3 in the USA).

## 3.1 Current Environment – Cybersecurity, Manufacturing & Government Stakeholders

*Question 1: What is your opinion regarding* ***the general need*** *for* *<u>cybersecurity</u>* *in the smart agriculture segment in Canada? (0-10 scale; 0 = no need, 10 = very large need)*

*Question 2: What would the* ***technical challenges /barriers to entry*** *be for the* *<u>cybersecurity</u>* *(if any)? (e.g. adoption, procurement, commercialization, processing, regulatory, other)?*

*Question 3: In your opinion, are there* ***any substitutes, emerging technologies, procedures or alternative products that could pose a threat*** *to the* *<u>cybersecurity</u>* technologies adoption in this sector*?*

### General Need

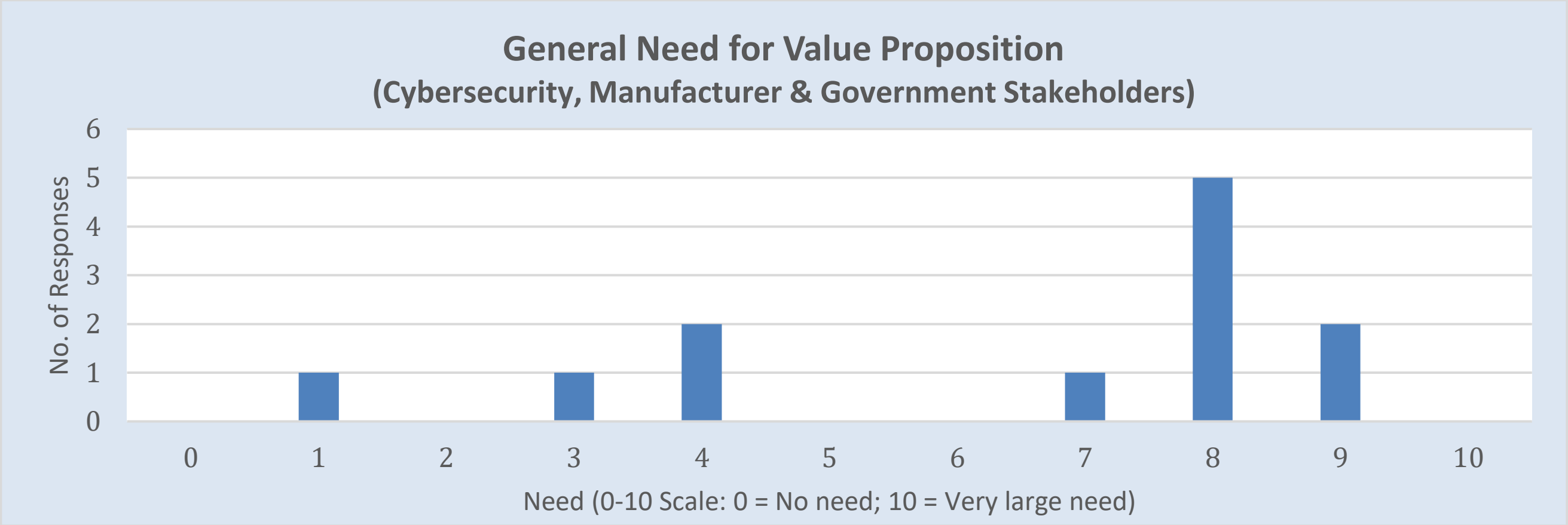

Cybersecurity, manufacturer & government stakeholders ranked **the general need** for cybersecurity systems for protecting smart and precision farming **as <u>moderate</u>**, with an **average score of <u>6.6/10</u>**, and the most frequent response being 8/10. That being said, there is a second, divergent group of opinion that places the general need much lower, with a rating between 1 to 4/10.[26]

### Technical Challenges/Barriers to Entry

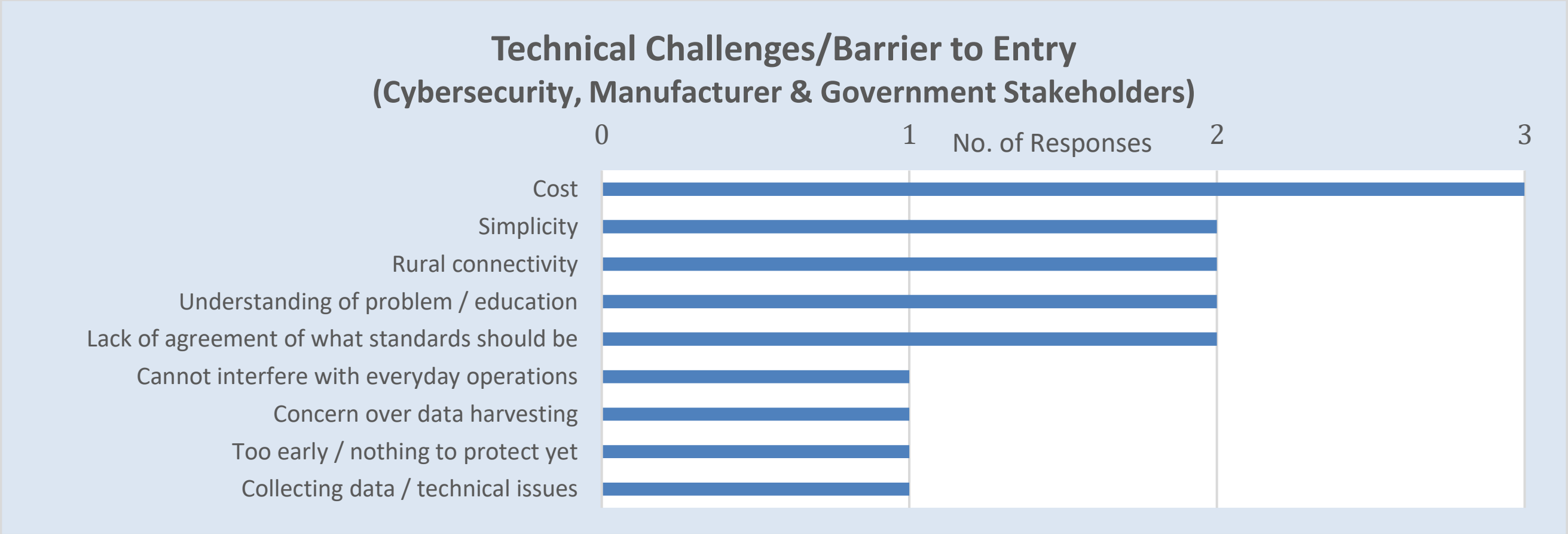

Cybersecurity, manufacturer & government stakeholders most frequently indicated that the single largest challenge/barrier is **<u>cost</u>**, followed by the simplicity of the proposed system, rural connectivity issues, understanding of the problem, and a lack of agreement as to what the standards should be.

[26] Some stakeholders provided two responses, based on their interpretation of the question. We have provided both numbers here, rather than an average.

## Substitutes, Emerging Technologies, Procedures or Alternative Products that Could Pose a Threat

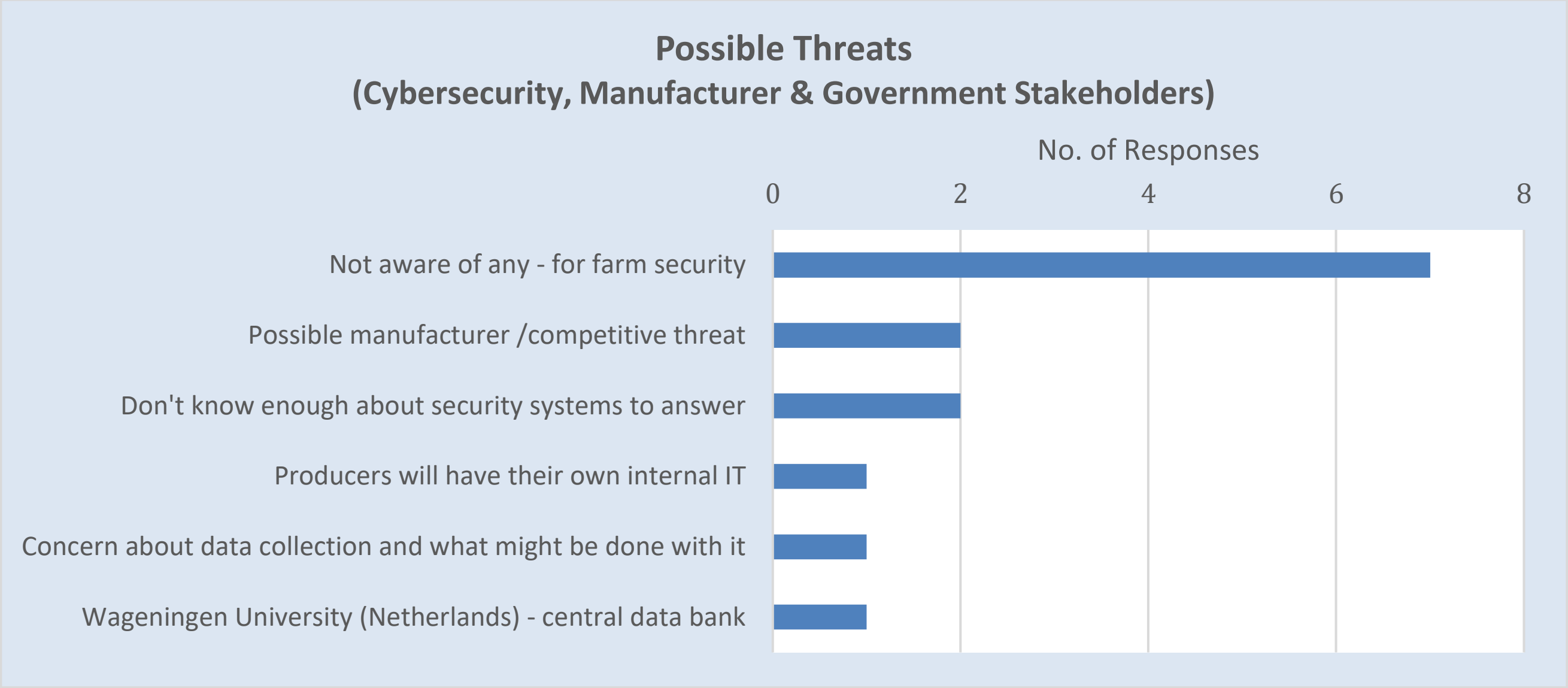

Cybersecurity, manufacturer & government stakeholders most frequently indicated that **they were not aware of any** substitutes, emerging technologies, procedures, or alternative products that could pose a threat to the proposed technology.

## 3.2 Current Environment – Digital Agriculture / Farming Stakeholders

*Question 4: **How significant an issue** is a cybersecurity to you, from a smart farming /agriculture perspective?*

*Question 5: Have you ever experienced a **cyberattack before?** If yes, what was the nature of the attack? How much damage did it result in? What was the nature of the damage? (financial loss, reputation, customer trust, etc.)*

*Question 6: Have you ever experienced **data breaches?** If yes, what were the consequences?*

*Question 7: **How satisfied are you with currently available solutions** (if any) in cybersecurity for protecting smart and precision farming/agriculture? (0-10 scale; 0 = very dissatisfied, 10 = very satisfied) Please explain why you rated the current solutions the way you did.*

### Significance of the Issue of Cybersecurity & Threat Hunting

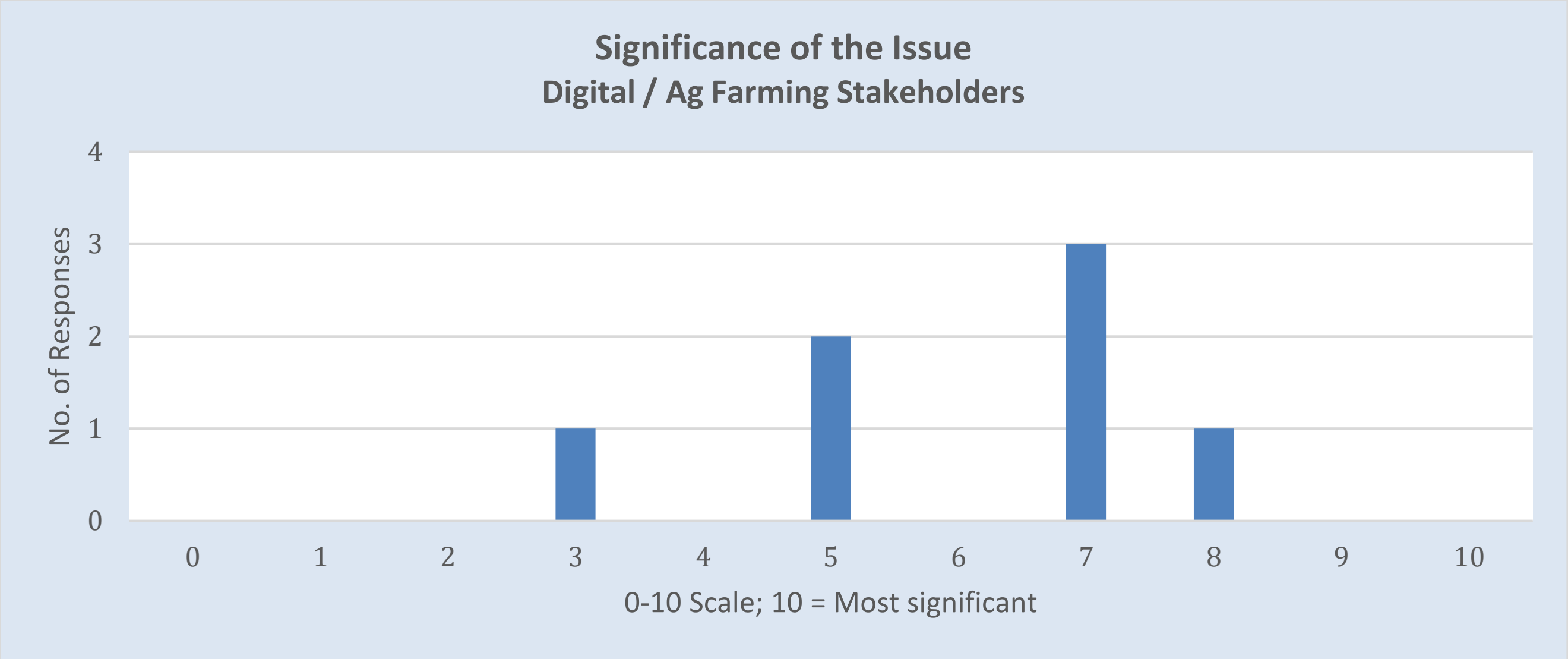

Digital agriculture / farming stakeholders indicated that cybersecurity is a **moderately significant issue**, with an average score of 6.1/10; the most frequent response being 7/10.

### Have You (or your Clients) Experienced a Cyberattack/Data Breach?

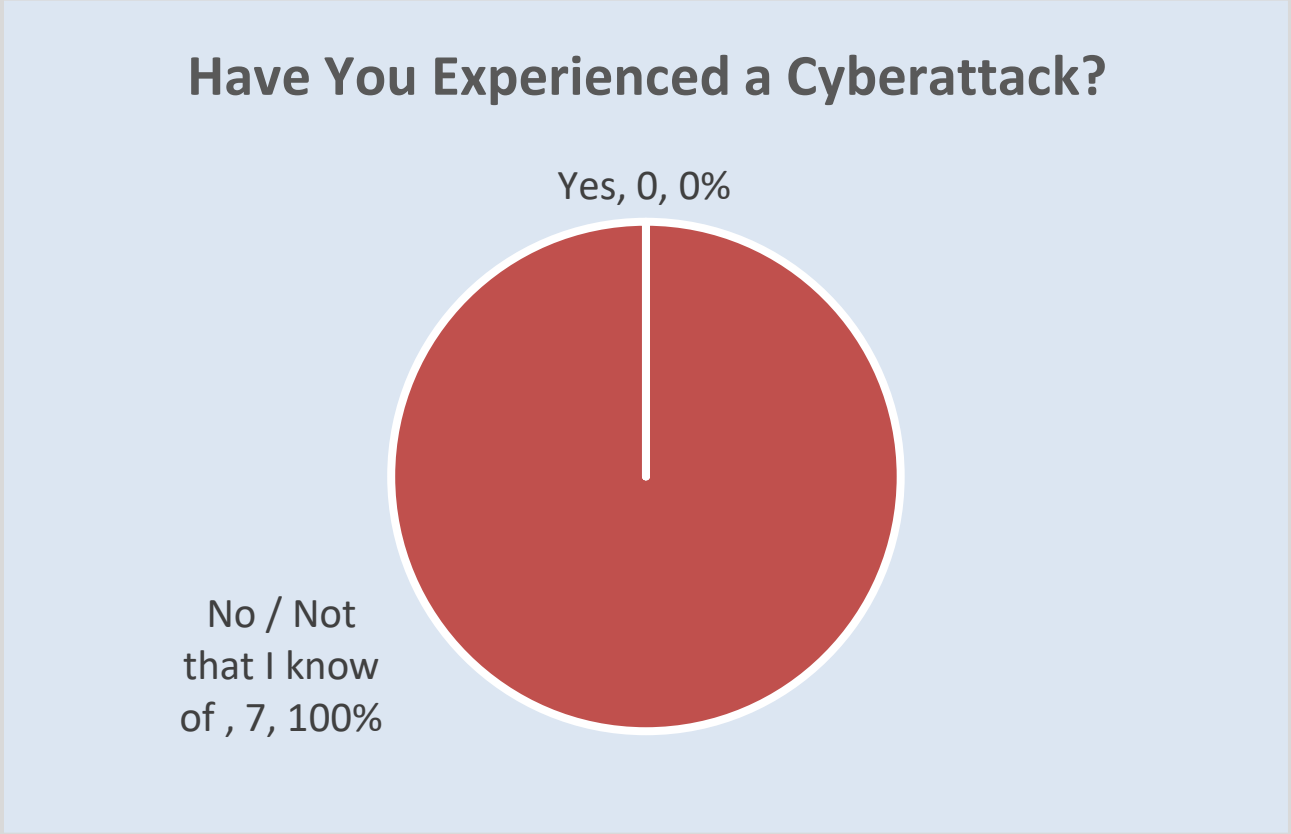

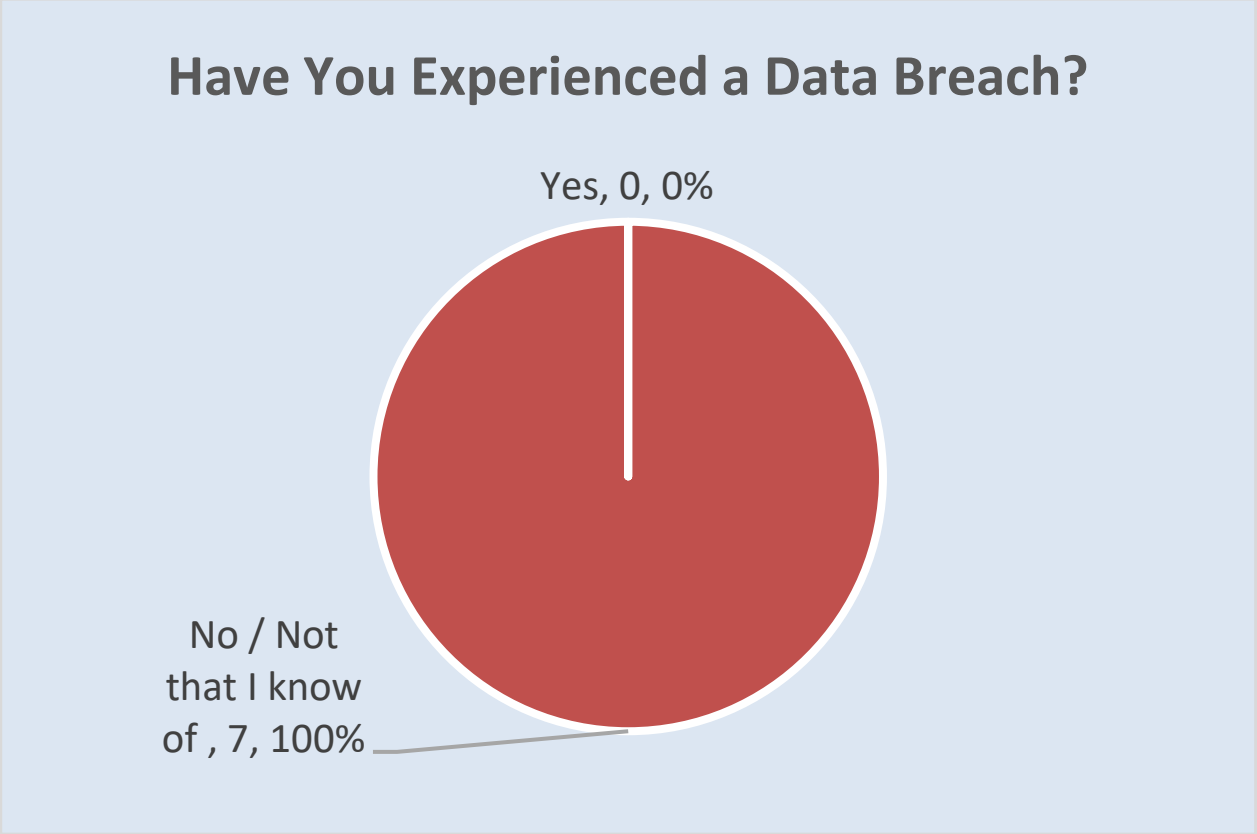

<u>**None**</u> of the digital agriculture/farming stakeholders interviewed has ever experienced either a cyberattack or a data breach, to the best of their recollection.

### Satisfaction with Current Solutions

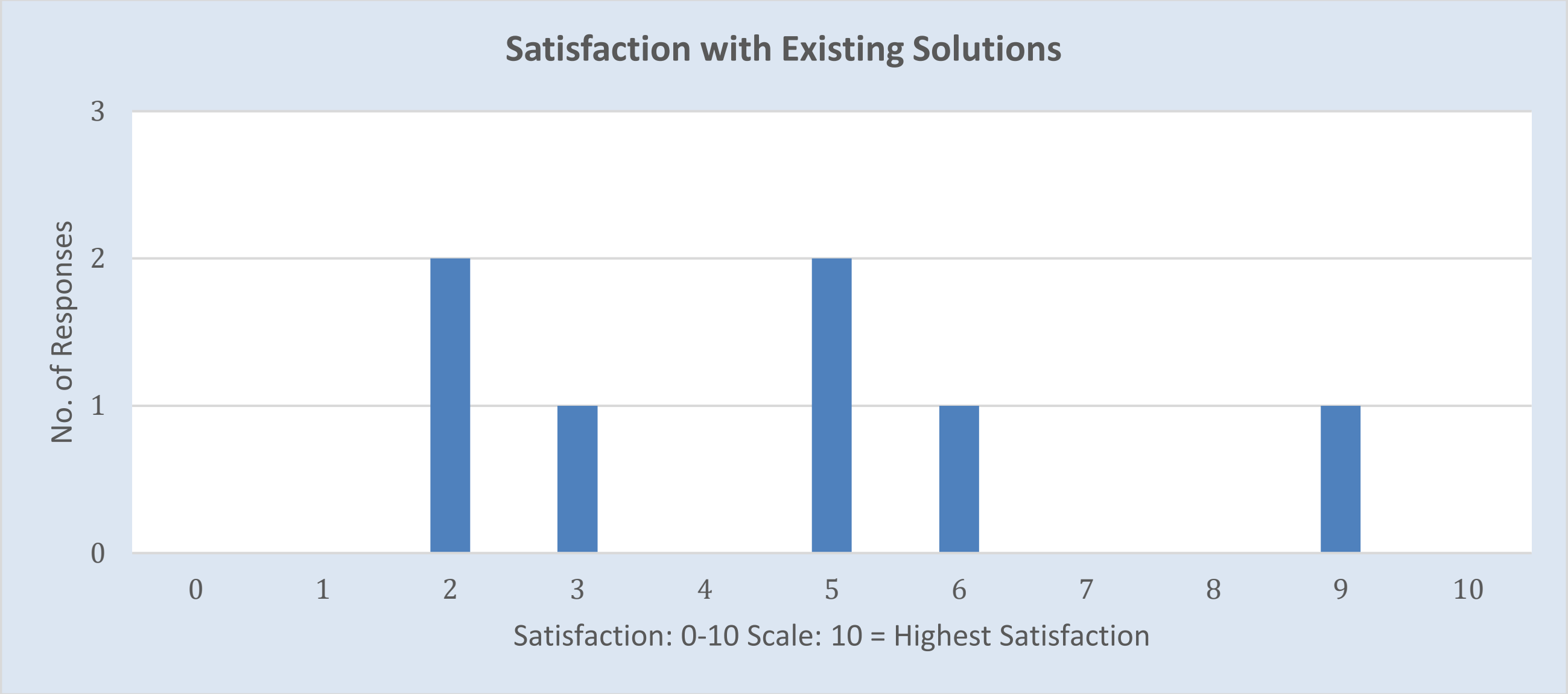

Digital agriculture / farming stakeholders are generally **not satisfied** with currently available solutions for cybersecurity in smart and precision farming/agriculture, providing an average score of **4.5/10**.

## 3.3 Current Environment – All Stakeholders

*Question 8: What do you think are the* ***missing components*** *(if any) in the existing monitoring systems for smart and precision farming?*

*Question 9: What do you* ***need to be included in the existing cybersecurity systems*** *for protecting smart and precision farming?*

### Missing Components in Existing Systems

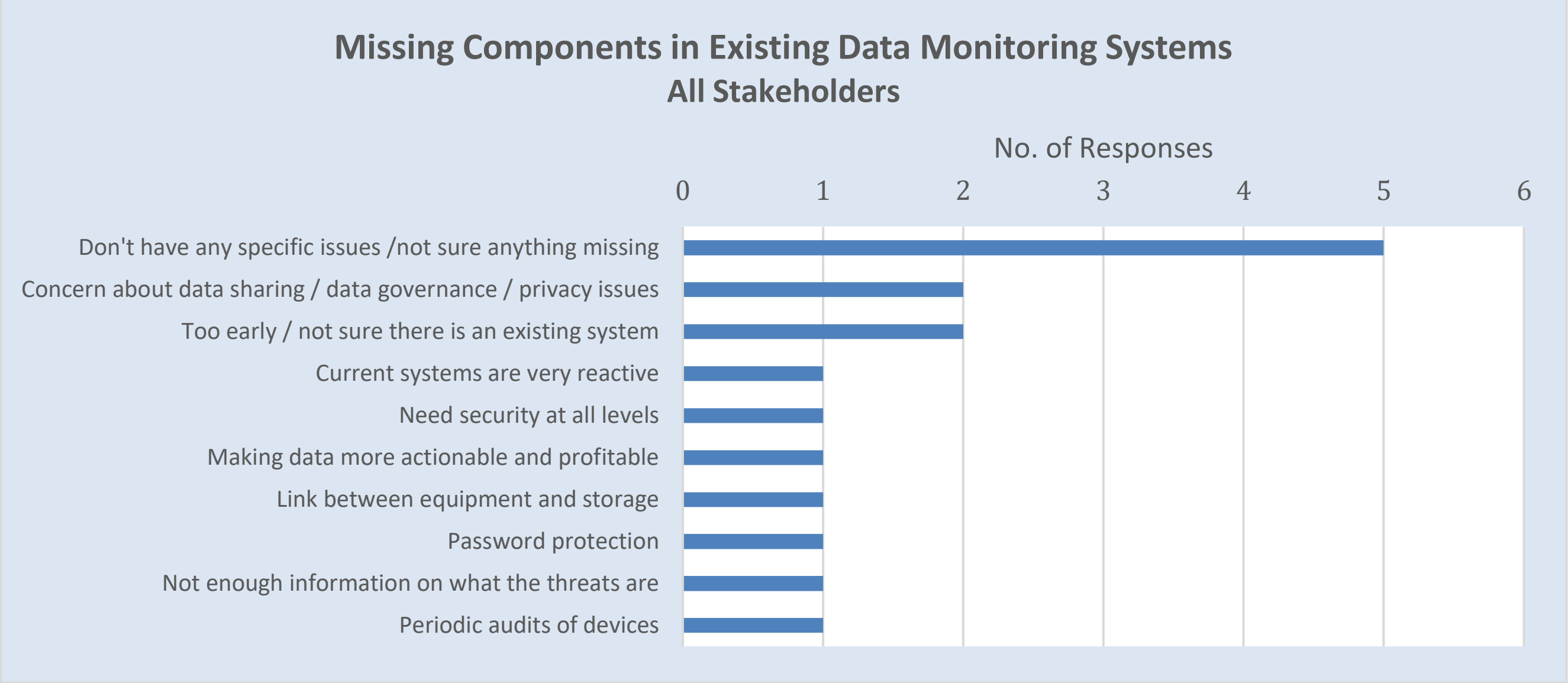

The most common response by all stakeholders to the issue of what is missing with existing systems for smart and precision farming is that **there are no outstanding issues they are aware of / they are not sure that anything is missing** – mentioned by about one-third of respondents.

### What Do You Need Included in Existing Systems?

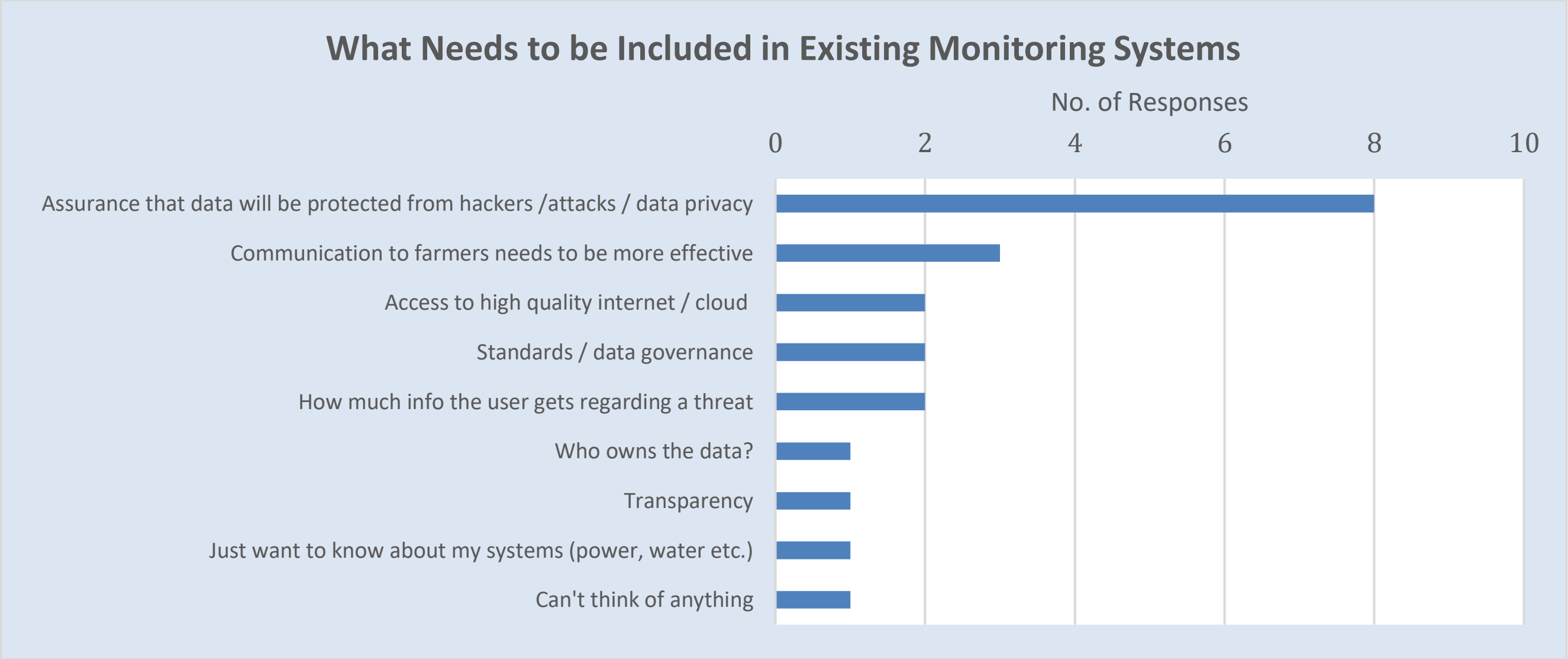

The most common response by all stakeholders to the issue of what needs to be included in existing systems for smart and precision farming is **assurance that data will be protected from hackers / attacks / data privacy.**

## 3.4 Overall Reaction to the Value Proposition

*Question 10: What is your **overall reaction** to the <u>cybersecurity systems for protecting smart and precision farming</u> as described as a solution for ensuring safe, secure and private smart farming? (Please state all positive and/or negative comments that come to mind. Please also rate it on a 0-10 scale, where 0 = very low-no interest/very low importance, and 10 = very high interest/very high importance)*

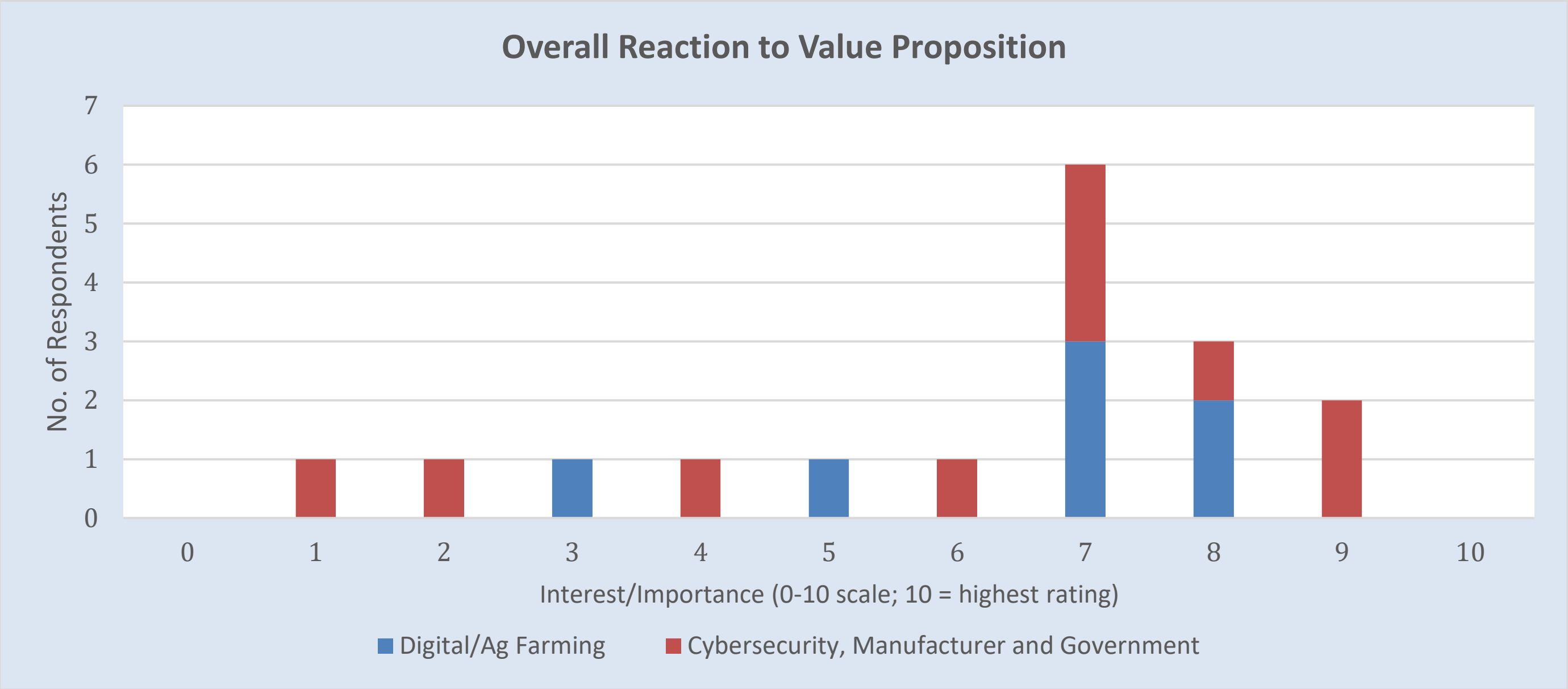

Overall, stakeholders felt that the cybersecurity systems for protecting smart and precision farming are **somewhat important, with an overall rating of <u>6.3/10</u>**. There was some degree of consensus in answer to this question, with the most frequent rating being 7/10, provided by 6 respondents – over one-third of the total.

## 3.5 Rating Proposed Advantages of Value Proposition

*Question 11: Please **rate the proposed advantages** of the <u>cybersecurity systems for protecting smart and precision farming value proposition</u>. (On a 0-10 scale: 0 is of no importance, 10 is very important). Please explain your ratings choices. (ask only the benefits that match the stakeholder)*

*A) Help smart farming system manufacturers to build **more secure and resilient systems** that can help farmers to detect and avoid cyberattacks;*
*B) The proposed privacy framework **eliminates the chance of data breach, liable to any compromise, loss of reputation, interruption of processes and destruction of equipment, and gaining improper competitor advantages**. This could significantly reduce cost and promote security of smart farming systems.*
*C) The proposed SIEM solution would **provide the required basis to conduct a data security audit of SF and PA networks**. Leveraging our collected data, farmers may be able achieve better control over their production process and control fellow of data in their businesses and further grow their production capacity through secure automation.*
*D) This project **advances knowledge and state-of-the-art in intrusion detection, intrusion analysis, and forensics examination** of smart and precision farming devices.*
*E) This project **provides a concrete view of the threat landscape in SF and PA** which paves the way for future Canadian startup companies to offer innovative solutions for securing SF systems, which is the expected long-term impact of this project.*
*F) Provides **government organizations and decision-makers with a thorough view of the cyberattack landscape and threats applicable to smart farming systems, in order to ensure that the appropriate policy and regulations** are in place.*
*G) This project is the first step towards **building a unique research center in North America**, focused on security and resiliency of smart and precision farming systems at the University of Guelph, contributing towards keeping it as the leading research university in Agri-tech sector of Ontario and Canada.*

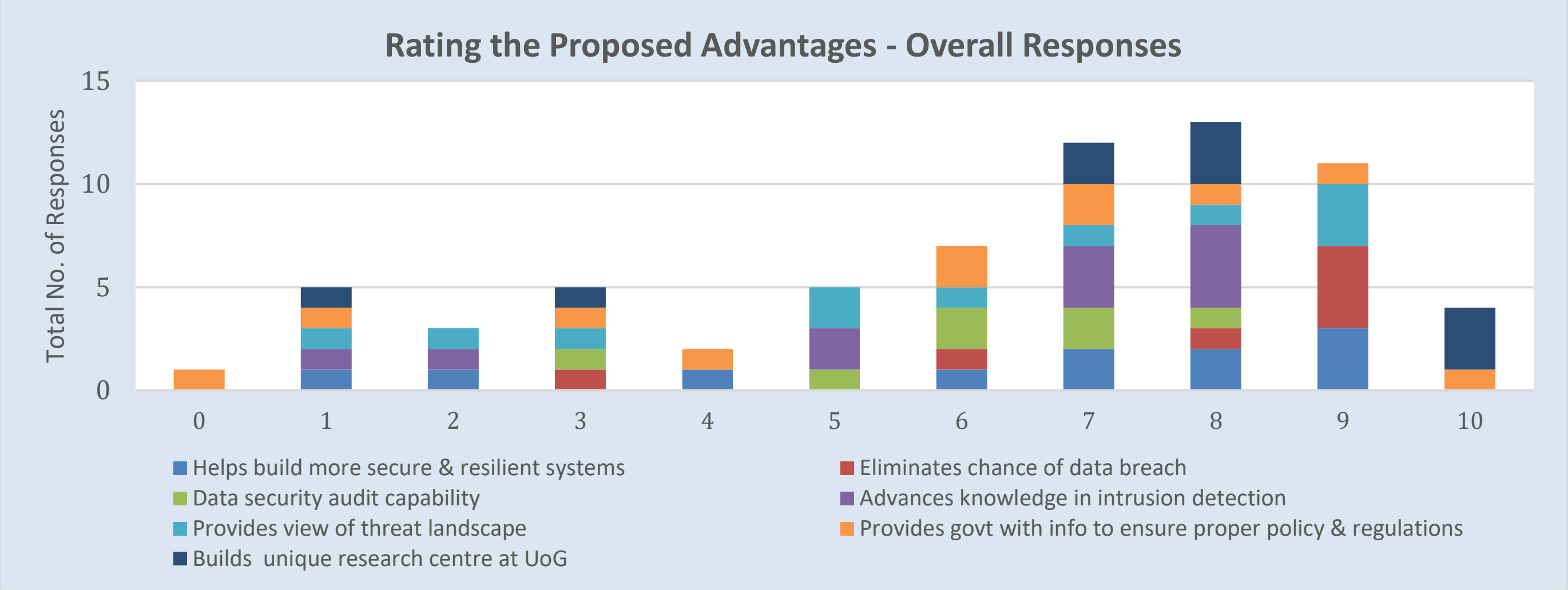

Overall, the total average for all scores on the 7 proposed advantages was **<u>6.3/10</u>** in terms of importance. The most frequently response was 8/10, provided 13 times, or .75 times for every respondent on average.

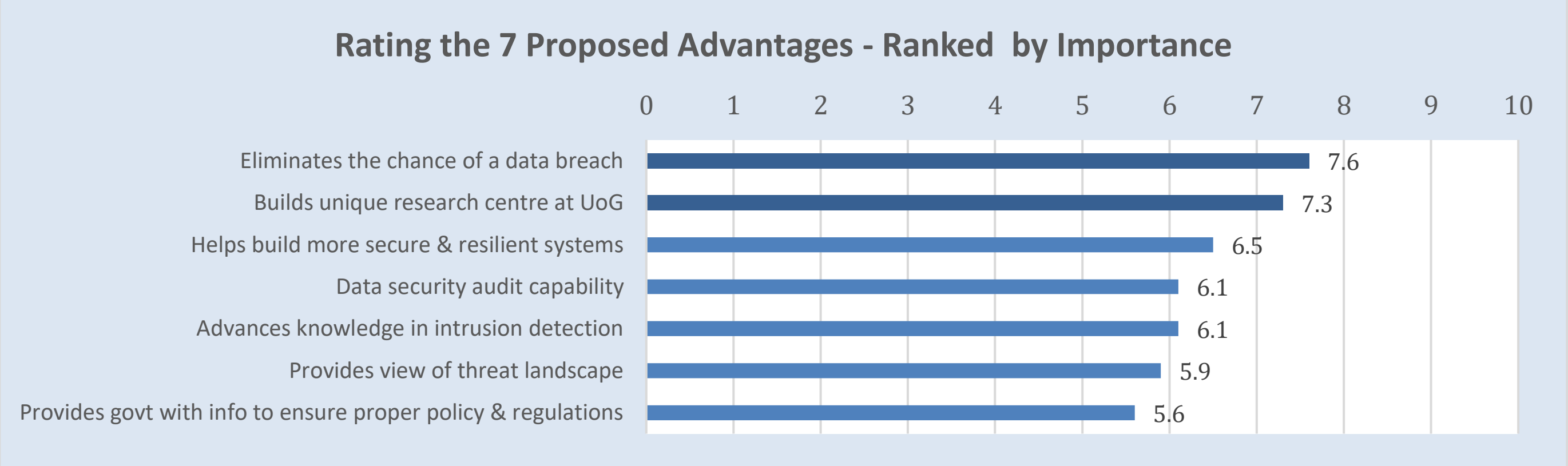

Overall, **the most important advantage** was eliminating the chance of a data breach (7.6/10).

# 4.0 CONCLUSIONS

Farmer and agri business adoption of smart and precision agriculture continues to grow, as discussed in the report. The scale and scope of complexity of the smart and precision technologies on the farm are increasing as IoT sensors are being incorporated throughout the whole operation. The amount of data being generated from farming operations and potentially shared on several platforms introduces new interdependencies on both hardware and software. The number of potential threat attack points continue to increase as the nodes on the smart farming network increase.

**The potential cost of disruption by a cyberattack to individual farming operations is growing in significance as the value dependency on software and hardware systems increases**. Farm machinery and mission critical equipment like dairy milking robots require the operating software to be working without disruption, and data protection and integrity are necessary. Get hacked, and you are out of business until it is fixed. The threat of cyberattacks is an important global issue as confirmed in the *U.S. Department of Homeland Security*[27] report. While the threat and potential harm of cyberattacks continues to grow, there are resources being directed to combat the blitz of viruses, malware, ransomware and other pests.

**There are no clear specialized "agriculture only" cybersecurity products/services or providers in the Canadian market but there is cyber protection being provided and developed**. Larger smart farming services providers such as *John Deere*, *Bayer – Climate Corporation*, *Cargill* and *Maple Leaf Foods* will be bundling cybersecurity features into their online farm management product offerings. The farm internet service providers will continue to provide some cybersecurity features. An important business tool for farmers is email – most are using free email services like *Gmail* and *Microsoft Outlook* where there are cybersecurity measures included like spam filters.

Farmers are also dependent on their **Smart phone suppliers – *Bell, Rogers,* and *Telus* to provide cybersecurity protection on their cellular coverage, which will be an important mode of data transfer on the farm.** The farmer use of mobile apps continues to grow as sensors and software become widespread – for example, watching satellite crop sensing imagery in products like *Fieldview* or *Granular*. Similarly, there are a growing number of digital protection companies for consumers like *Kaspersky* that provide easy to use and inexpensive tools that farmers will have to learn how to use to remove malware.

**Small and medium-sized agriculture and food businesses will need to develop relationships with cybersecurity firms to protect their farmer facing systems and servers**. Website hosting companies with agriculture and food clients are providing some cybersecurity services. Large technology suppliers like Amazon Web Services (AWS) will continue to grow their share of the agriculture industry due to the competitive scale and scope of their services and cyber protection infrastructure.

The market research results presented in this report on the product-market fit of the proposed cybersecurity services **seem reasonable and consistent with past experience**.

## Technology Sector Specialist Conclusions

The potential threat of costly disruption to farming operations and agri businesses from cyberattacks will continue to grow as adoption and integration of information technologies become widespread.

Farming operations have been slow to adopt and integrate new technologies for several reasons – they are small rural businesses and often do not have internal IT staff or expertise. It often takes a cyberattack to disrupt a business or farm to create demand for a product/service like this. As the respondents to the survey have indicated, there is a perception that there have not been many serious attacks with malware or ransomware. Asking network administrators of agriculture and food companies and you would get a different answer – websites and networks are under constant attack and when hacked and down, companies quietly try to restore things without customers noticing. There have been very few publicly announced incidents like the *City of Stratford* or *Meridian Automotive* so most people are uninformed. Nothing will get your attention as a business owner like being informed your technology is down because of ransomware. **There is a customer market need, but it is just emerging**.

[27] https://www.dhs.gov/sites/default/files/publications/2018%20AEP_Threats_to_Precision_Agriculture.pdf

The cost of a disruptive cyberattack will be significant to farmers and agri business so there will be a business opportunity for security providers, the major question will be – is there enough value creation for being agri food specialists to carve out a profitable and sustainable niche in the market? Based on past experience with farmer and SME agri businesses, the adoption of new technology offerings is very slow. **A company commercializing a cybersecurity monitoring & threat hunting system for protecting smart and precision farming will have to overcome several technology and marketing challenges in their "go to market strategy" in order to be financially sustainable**.

1. There will be **non-traditional technology competitors to an agri food specialized cybersecurity services company and they will be formidable**, Amazon, Microsoft, IBM, Google and specialized industrial and consumer IT firms will have SME and consumer offerings.
2. A new organization could also **research and educate the industry on overlapping technologies such as blockchain and crypto currencies which are designed to secure financial and transparency**. *Fred Wall* at *Farm Credit Canada* has been taking on a leadership role with issues like Data Transparency and innovation to help improve the competitiveness in Canada.
3. One implication of the globalization of cybersecurity services will be that **it is unlikely that many of the servers and technology will be based in Canada**, so there may be reasons for a "Made in Canada" services offering – e.g. the Cannabis industry needs to avoid US legal restrictions.
4. There is **a need for a significant investment in stakeholder education and high-level monitoring of cyber threat at the provincial and federal government level to ensure our agriculture and food sector is doing everything it can from a policy perspective.** Organizations such as *OMAFRA* and *AgriCorp* likely are implementing cybersecurity procedures but would benefit greatly from agri specific expertise to help overcome the risk of rural infrastructure-based attacks.

# Bibliography

AgFunder**, AgFunder Agri-food Tech Investing Report – 2018,** https://agfunder.com/research/agri-food-tech-investing-report-2018/

Allied Market Research, **Global IOT in Agriculture Market Expected to Garner $48.71 Bn by 2025 at 14.7% CAGR, Says AMR** , www.globenewswire.com/news-release/2019/07/22/1885778/0/en/Global-IOT-in-Agriculture-Market-Expected-to-Garner-48-71-Bn-by-2025-at-14-7-CAGR-Says-AMR.html

ArgoCares, **What is the difference between precision, digital and smart farming?** , www.agrocares.com/en/news/precision-digital-smart-farming/

CBInsights, **AI, Robotics, And The Future Of Precision Agriculture**, July, 2017, www.cbinsights.com/research/ai-robotics-agriculture-tech-startups-future/

CEMA, **Digital Farming: what does it really mean?** February, 2017, www.cema-agri.org/images/publications/position-papers/CEMA_Digital_Farming_-_Agriculture_4.0__13_02_2017_0.pdf

Ghaffarzadeh, Dr Khasha, **Agricultural Robots and Drones 2018-2038: Technologies, Markets and Players**, www.idtechex.com/en/research-report/agricultural-robots-and-drones-2018-2038-technologies-markets-and-players/578

Grandview Research, **Precision Farming Market Worth $10.23 Billion By 2025 | CAGR: 14.2%** , May, 2019, www.grandviewresearch.com/press-release/global-precision-farming-market

IoT Agenda, **IoT cybersecurity: A major concern for agriculture**, https://internetofthingsagenda.techtarget.com/blog/IoT-Agenda/IoT-cybersecurity-A-major-concern-for-agriculture

IoT Agenda, **IoT security (internet of things security)**, https://internetofthingsagenda.techtarget.com/definition/IoT-security-Internet-of-Things-security

ISA, **Cybersecurity for the Agricultural Industry,** May, 2019, www.e-isa.com/blog/cybersecurity-for-agriculture.html

Jahn Research Group, **Cyber Risk and Security Implications in Smart Agriculture and Food Systems,** January, 2019, https://jahnresearchgroup.cals.wisc.edu/wp-content/uploads/sites/223/2019/01/Agricultural-Cyber-Risk-and-Security.pdf

Kukutai, Arama, **Driving new revenue sources through a digital ag revolution,** August, 2016, https://techcrunch.com/2016/08/08/driving-new-revenue-sources-through-a-digital-ag-revolution/

MarketsAndMarkets , **AI in Agriculture Market by Technology (Machine Learning, Computer Vision, Predictive Analytics), Offering, Application (Precision Farming, Drone Analytics, Agriculture Robots, Livestock Monitoring), Offering, and Geography - Global Forecast to 2025** , www.marketsandmarkets.com/Market-Reports/ai-in-agriculture-market-159957009.html

Market Stats Report, **Artificial Intelligence in Agriculture Market: Global Market Size, Trends, Competitive, Historical & Forecast Analysis, 2019-2025** , https://marketstatsreport.com/artificial-intelligence-in-agriculture-market-size/#REPORT-DESCRIPTION

Miller, Carol, **Precision** Agriculture **in Specialty Crops: Labor, Cost Savings are Key,** October, 2018, www.precisionag.com/in-field-technologies/precision-agriculture-in-specialty-crops-labor-cost-savings-are-key/

Morgan, Steve, **Global Cybersecurity Spending Predicted To Exceed $1 Trillion From 2017-2021**, June, 2019, https://cybersecurityventures.com/cybersecurity-market-report/

Neuman, Daniel, **Top Six Digital Transformation Trends In Agriculture,** May, 2018, www.forbes.com/sites/danielnewman/2018/05/14/top-six-digital-transformation-trends-in-agriculture/#3bffee41ed2e

O'Connor, Michael, **Precision Agriculture: Opportunities and Challenges**, www.nap.edu/read/13292/chapter/20

Public-Private Analytic Exchange Program-DHS, **Threats to Precision Agriculture**, www.dhs.gov/sites/default/files/publications/2018%20AEP_Threats_to_Precision_Agriculture.pdf

Ruffalo, Alan, **Artificial Intelligence in Agriculture Market Key Players, Industry Overview, Supply Chain and Analysis to 2019-2025**, March, 2019, http://bmrcnews.com/artificial-intelligence-in-agriculture-market-key-players-industry-overview-supply-chain-and-analysis-to-2019-2025/

Sciforce, **Smart Farming: The Future of Agriculture**, February, 2019, www.iotforall.com/smart-farming-future-of-agriculture/

Sela, Guy, **Why Are 570 Million Farmers Not Yet Using Agricultural Apps?,** September, 2018, www.precisionag.com/digital-farming/analytics/why-are-570-million-farmers-not-yet-using-agricultural-apps/

Shahbandeh, M, **Forecast market value of smart agriculture worldwide in 2017 and 2022 (in billion U.S. dollars),** January, 2019, www.statista.com/statistics/720062/market-value-smart-agriculture-worldwide/

Steele, Dale, **Analysis of Precision Agriculture Adoption & Barriers in western Canada,** April, 2017, www.realagriculture.com/wp-content/uploads/2017/04/Final-Report-Analysis-of-Precision-Agriculture-Adoption-and-Barriers-in-western-Canada-April-2017.pdf

TheInternetofallthings, **Cost of US IoT cybersecurity plan over $35 million**, September, 2019 https://www.theinternetofallthings.com/cost-of-us-iot-cybersecurity-plan-over-35-million

Vanderlinde, Mark, **NEW TRENDS IN AGRICULTURE TECHNOLOGY**, www.cenex.com/about/cenex-information/cenexperts-blog-page/agriculture-and-farming/ag-tech-trends

Variant Market Research, **Smart Agriculture Market Overview,** www.variantmarketresearch.com/report-categories/information-communication-technology/smart-agriculture-market